# Voltage Estimation for Diode-Clamped MMC Using Compensated State-Space Model

N. Tashakor, Y. Zhang, S. Banana, F. Blaabjerg, S. Goetz

*Abstract*—**Modular multilevel converters are well known in the energy sector and have significant potential in others, such as electro-mobility. Generally, their stable operation is at the expense of numerous sensors, communication burden, and complicated balancing strategies that challenge its expansion to cost-driven applications. Hence, the introduction of a sensorless voltage-balancing strategy with a simple controller is an attractive objective. Diode-clamped MMCs offer a simple and yet effective solution by providing a unidirectional balancing path between two modules through a diode, and ideally the modulation technique should compensate the lack of bidirectional energy transfer; hence open-loop operation is possible. Although open-loop operation is desirable to reduce costs, good knowledge of the modules' voltages for system monitoring and protection functions still appears mandatory. This paper develops a compensated state-space model for diode-clamped MMCs and integrates it with an optimal estimator to reliably estimate the voltage of all modules without any direct measurement at module terminals. Additionally, the model considers the effect of the diode-clamped branches and their balancing effect, resulting in 30% to 50% reduction in estimation error compared to the conventional model. Simulations and experiments further confirm the provided analysis, where the estimator achieves above 97.5% accuracy.**

## I. Introduction

Modular multilevel converters (MMCs) are a well-known solution in many high-voltage applications [1]. The main practical advantages compared to other converters are an excellent harmonic performance through quantized voltage levels, easier scalability, flexibility due to their modularity, and fault ride-through capability [2, 3]. Such features render MMCs particularly valuable in medium- to high-voltage applications such as high voltage direct current (HVDC) transmission, distributed generation as well as the basis for energy storage systems and solid-state transformers [4, 5]. However, despite numerous advantages, there are still caveats that need to be addressed, of which balancing and voltage monitoring are among the most critical ones [6, 7].

MMCs with diode-clamped modules can be considered as the simplest among self-balancing topologies and presumably the only ones that do not require extra active switches [8-11]. In these topologies, extra diode-clamped-paths can create parallel connections neighboring modules [12, 13]. Fig. 1 presents the simplest form of a diode-clamped topology, with the clamping path marked in green. Clamping path forms a balancing backbone, which can even ensure balancing in hardware for high safety-integrity levels. However, the diode can only conduct in one direction and hence the provided balancing path will be unidirectional [14, 15]. Although more complex clamping topologies to create a physical bidirectional balancing path exist [9, 12, 16, 17], the required additional diodes, transformers, or even transistors defeat the main incentive behind using diode-clamped topologies to reduce cost and complexity [18].

A more practical approach can ensure that the needed direction of power balancing naturally complies with the diode conduction direction [19]. Such natural balancing direction is closely related to ordering modules and can easily be achieved by manipulating the modulation reference of the individual modules, such as level-adjustment phase-shifted carrier modulation (LAPSC) [16, 20].

Although there are other sensorless and/or open-loop balancing methods for diode-clamped MMCs, knowing the exact voltage of each module can be necessary for monitoring and protection functionalities. Additionally, as we will discuss in IV-B, the state-of-the-art estimation techniques for conventional MMC topologies, such as [21-26], can lead to sub-optimal results mainly since they do not consider the effect of the clamping branch as well as the balancing control in the MMCs.

Therefore, this paper develops a compensated state-space model for diode-clamped MMCs that takes into account the effect of the clamping branches as well as balancing. The proposed model in combination with an appropriate estimator can provide a stable and accurate estimation of the modules' voltages. Additionally, the developed method is compatible with balancing strategies, such as LAPSC, and can offer a more stable estimation under imbalance and does not require any direct measurement at module level, which reduces the communication and sensing costs considerably.

In the following, Section II provides a brief introduction into the diode-clamped topology. Section III studies the principles of balancing, and Section IV develops the necessary state-space model and estimator. Simulation and experimental results in Section V verify the performance of the proposed technique. Finally, Section VI concludes the paper.

## II. Diode-Clamped MMC

Fig. 1 shows the single-phase topology of the diode-clamped MMC containing two identical arms. The three-phase structure can also be developed based on the single-phase topology.

### *A. Operation of the Clamping Circuit*

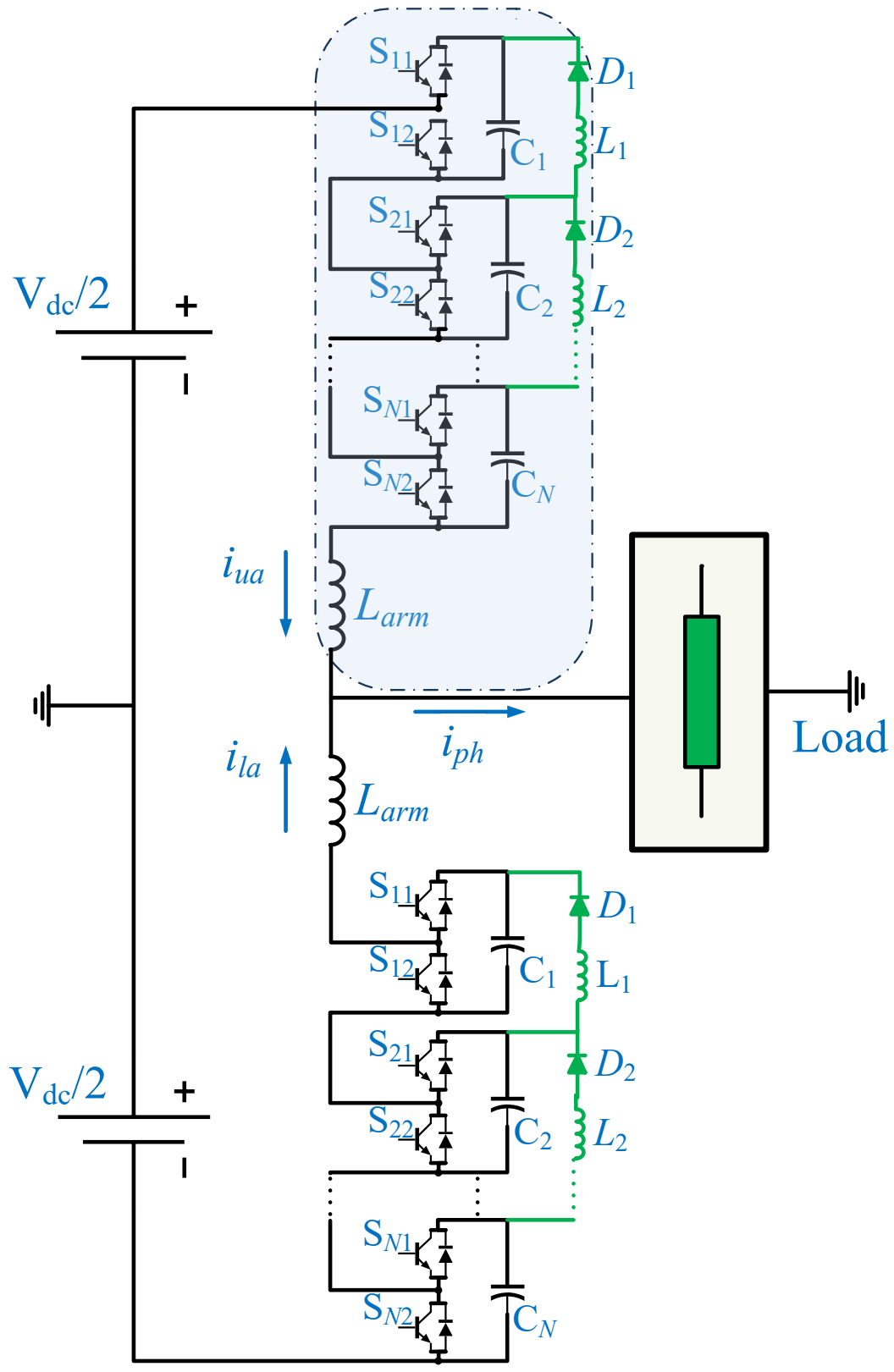

Fig. 1. Single-phase topology of a diode-clamped MMC

Each arm contains $N$ modules, but only $(N-1)$ clamping circuits are necessary. Each clamping circuit consists of a diode and an inductor in series. The very small inductor—as low as few microhenries and therefore with negligible core size [27, 28]—limits the maximum balancing current, in the case of high voltage differences between the modules. The voltage across the clamping branch ($v_{b_i}$) includes the diode/switch forward voltage drop ($V_{fd}$) and the inductance voltage ($v_{L_i}$) as well as the voltage across the parasitic resistance of semiconductors, capacitors, and inductors ($v_{R_{sum}}$) per

$$v_{b_i} = 2V_{fd} + v_{L_i} + v_{R_{sum}}, \tag{1}$$

where $2V_{fd}$ is the voltage drop across the clamping diode $D_i$ as well as the main switch $S_{(i+1)2}$ or reverse diode of $S_{(i+1)1}$, assuming an identical voltage drop for the diodes and switches ($V_{diode} \approx V_{switch}$).

Depending on the control signal of the $(i+1)^{th}$ module as well as the module voltages, four possibilities can be imagined. When $S_{(i+1)1}$: off and $S_{(i+1)2}$:on, the voltage across the clamping branch is $v_{b_i} = V_{c_{i+1}} - V_{c_i}$. If $V_{c_{i+1}} \leq V_{c_i} + 2V_{fd}$, the branch is open as Fig. 2(a) demonstrates (Mode 1), but if $V_{c_{i+1}} > V_{c_i} + 2V_{fd}$, a balancing current flows from $C_{i+1}$ to $C_i$ as Fig. 2(b) illustrates (Mode 2). When $S_{(i+1)1}$: on and $S_{(i+1)2}$:off, the diode is reverse-biased, and, as Fig. 2(c) depicts (Mode 3), the balancing current decays to zero. Assuming negligible resistive elements, the clamping inductor determines the decay rate given as

$$\frac{di}{dt} = \frac{-(V_{C_i}+2V_{fd})}{L_i}. \tag{2}$$

Large imbalances between two modules can require multiple sequences of Modes 2 and 3, until the voltage difference of the modules is cleared ($V_{c_{i+1}} - V_{c_i} \leq 2V_{fd}$). Fig. 3 illustrates an intuitive representation of this situation. The analysis can be extended to all of the modules in an arm following

$$V_{c_1} + 2V_{fd} \geq ... \geq V_{c_{(N-1)}} + 2V_{fd} \geq V_{c_N}. \tag{3}$$

*B. Circuit analysis*

The equivalent electrical circuit when $S_{(i+1)2}$ is on (forward bias) and $V_{c_{i+1}} > V_{c_i} + 2V_{fd}$ is a second-order RLC circuit [12, 20]. Based on Kirchhoff's voltage law and after some manipulations a second-order differential equation can be derived as

$$\frac{d^2 i_D(t)}{dt^2} + \frac{R_{sum1}}{L_i}\frac{di_D(t)}{dt} + \frac{1}{L_iC_e} i_D(t) = 0. \tag{4}$$

The equivalent voltage is $V_e = V_{diff} = V_{c_{i+1}} - V_{c_i}$, and the equivalent capacitance $C_e = 0.5C_i = 0.5C_{i+1}$.

Applying Laplace transformation and solving it results in

$$P_{1,2} = -\frac{R_{sum1}}{2L} \pm \sqrt{\frac{{R_{sum1}}^2}{4L_i^2} - \frac{1}{L_iC_e}}. \tag{5}$$

The equivalent resistance $R_{sum1}$ is comparably small; hence the current will be a damped oscillation given by

$$i_D(t) = \frac{V_{diff}}{\sqrt{\frac{L_i}{C_e} - \frac{{R_{sum1}}^2}{4}}} e^{-\alpha t} \sin \omega_d t \tag{6}$$

where the damping factor $\alpha = \frac{R_{sum1}}{2L_i}$, and the frequency of the oscillation is $\omega_d = \sqrt{\frac{1}{L_iC_e} - \frac{{R_{sum1}}^2}{4L_i^2}}$.

If the maximum permissible voltage difference between modules is $V_{diff_max}$, the maximum peak diode current ($I_{Pmax}$) follows

$$I_{Pmax} \leq \frac{V_{diff_max}}{\sqrt{\frac{L_i}{C_e} - \frac{{R_{sum1}}^2}{4}}}. \tag{7}$$

Therefore, solving for $L_i$, as long as the inductor follows

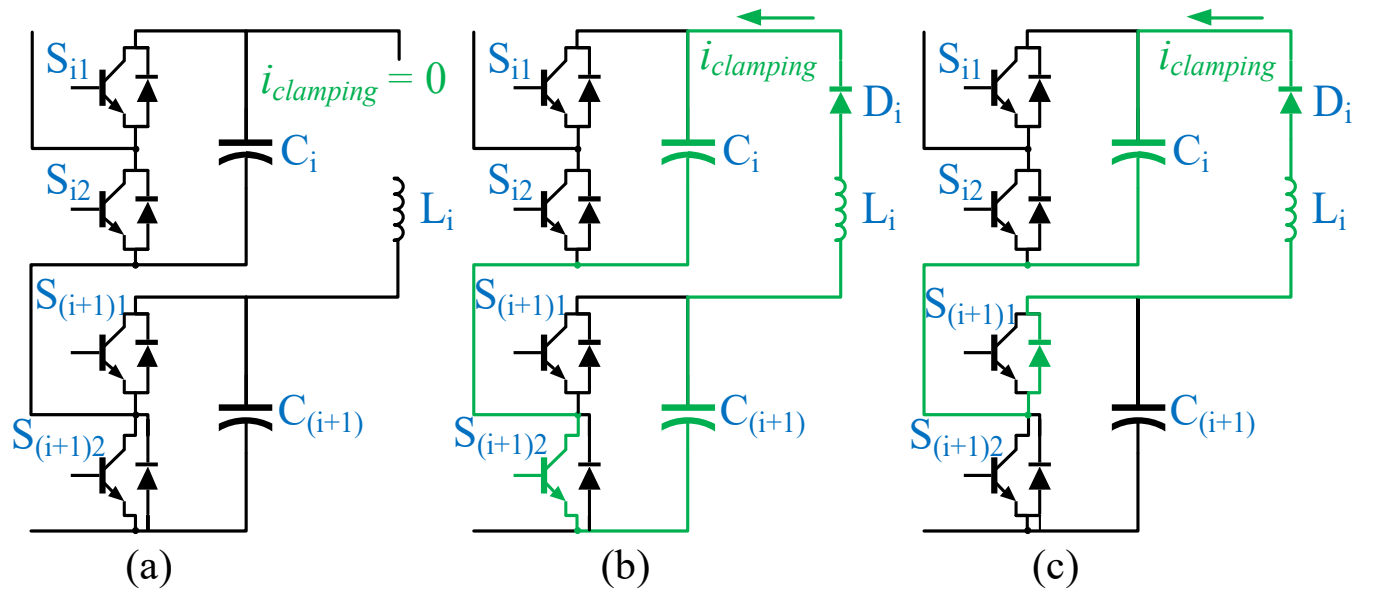

Fig. 2. Different operation modes of the diode-clamped modules: (a)$V_{c_{i+1}} \leq V_{c_i} + 2V_{fd}$ (Mode 1); (b)$V_{c_{i+1}} > V_{c_i} + 2V_{fd}$ (Mode 2); (c)$S_{(i+1)2}$turns off (Mode 3)

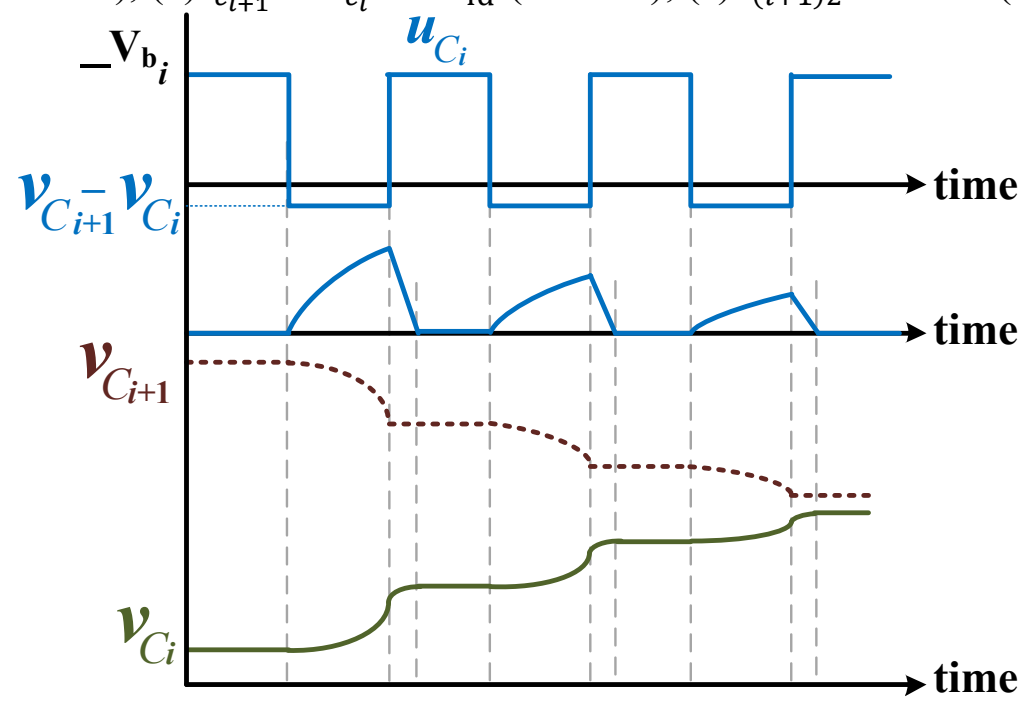

Fig. 3. Balancing process for two modules with diode-clamped topology

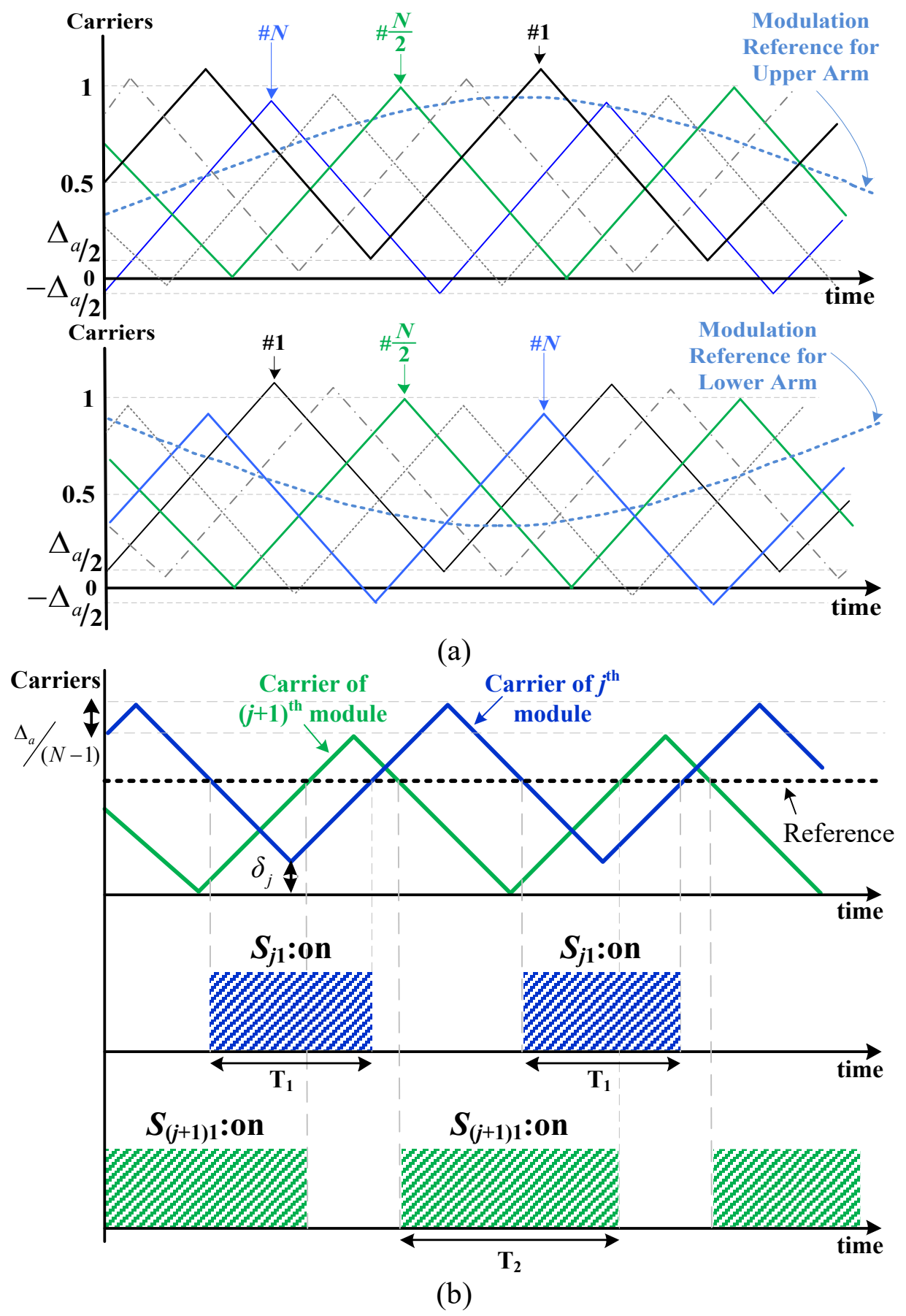

Fig. 4. Intuitive representation of the LAPSC (a) carrier placement in upper and lower arms; (b) effect of level-adjustment on the pulse-width

$$L_i \geq \left(\frac{{R_{\mathrm{sum1}}}^2}{4} + \frac{{V_{\mathrm{diff_max}}}^2}{{I_{\mathrm{P_max}}}^2}\right) C_e, \tag{8}$$

the current of diode $D_i$ is within its rating [20]. Additionally, it is possible to reduce the current rating of the diode by increasing the size of the inductor. However, a larger inductor value reduces the speed of the balancing [28].

## III. Phase-Shifted Carrier Modulation (PSC)

The conventional phase-shifted carrier (PSC) modulation compares a reference waveform (modulation index) with carriers that are phase-shifted with respect to each other. In PSC modulation, each carrier uniquely corresponds to one module in the arm and the phase-shift between two successive carriers is fixed to $\frac{2\pi}{N}$. With ideal conditions, PSC should reach a stable operating point [6]. However, the system gradually diverges from the intended operation point if no balancing mechanism exist [29].

A diode-clamped circuit can ensure that condition (3) is always maintained, but the voltages only converge if the balancing current flows at all times from lower modules in the arm to the higher ones, i.e., from bottom of the arm to the top. Although various balancing algorithms expose different behaviors and (dis)advantages, almost all of them ensure a correct balancing direction through manipulating the effective conduction time of the modules. Liu et al. implemented a similar method by actively manipulating the conduction time of the top module [20]. Zheng et al. introduced a time delay between the switching instances [16]. Level-adjusted phase-shifted carrier (LAPSC) can ensure the correct balancing direction through a small vertical adjustment to the normally in-line phase-shifted carriers (see Section III-A) [27]. Even, closed-loop methods, such as cell-sorting, implement a similar measure. Hence, the following analysis focuses on the general effect of balancing algorithms.

### A. Study of Balancing Method

Pulse-width manipulation is normally achieved by introducing a small offset to the modulation reference. The counter part of introducing an offset in modulation reference is a vertical level-adjustment in the corresponding carriers, as Fig. 4 intuitively depicts. A positive vertical level-adjustment (for the $i^{\text{th}}$ carrier, $\delta_i > 0$) reduces the average duration that the module is connected in series and negative level-adjustment ($\delta_i < 0$) increases the average duration of the series intervals.

Regardless of the balancing method, since the arm average current is positive, shorter durations with series connection reduce the module charge and vice versa. Subsequently, a zero adjustment (in the carriers) or offset (in the individual modulation indices) leads to all modules having the same average series connection duration. If $\delta_j$ represents the introduced positive vertical-adjustment of $j^{\text{th}}$ carrier (or a negative offset in the modulation reference), $\delta_1 \geq \delta_2 \geq \cdots \geq \delta_N$ ensures a bottom to top balancing direction at all times. Without lack of generalization, considering the upper arm, the effective modulation index ($m_j$) for the $j^{\text{th}}$ module with $\delta_{u,j}$ is

$$m_{u,j} = \frac{1 - m_a \sin(\omega t)}{2} - \delta_{u,j}, \tag{9}$$

where $m_a$ is the normalized amplitude of the phase $a$.

The effective modulation index $\left(m_u(t)\right)$ for the complete arm is the average of all the individual $m_j$ values as follows

$$m_u(t) = \frac{1 - m_a \sin(\omega t)}{2} - \frac{\sum_j^N \delta_{u,j}}{N}. \tag{10}$$

Defining the adjustments for the $j^{\text{th}}$ module in the upper arm per

$$\delta_{u,j} = \Delta_a \left(\frac{1}{2} - \frac{j-1}{N-1}\right) \tag{11}$$

results in a zero average for the term corresponding to the adjustment in (10), and hence an unchanged effective voltage of the arm. $\Delta_{\mathrm{a}}$ in (11) is the total adjustment between the first and last carriers of the arm.

Since the balancing direction in the lower arm must also be from bottom to top, $\delta_{l,j}$ should be equal to $\delta_{u,j}$ following (11). However, the carrier orders (phase shifts) in the upper and lower arm should be mirrored from the point that they are connected to each other, to ensure that at each instant the effects of the balancing effort (e.g., level-adjustments) on the phase $a$ are zero. Hence, the vectors of phase-shifts for the upper and lower arm should be

$$\boldsymbol{\phi}_{\mathrm{u}} = \left[0. \frac{2\pi}{N}, \frac{4\pi}{N}. \ldots. \frac{2\pi(N-1)}{N}\right]^T \tag{12}$$

and

$$\boldsymbol{\phi}_{\mathrm{l}} = \left[\frac{2\pi(N-1)}{N}, \frac{2\pi(N-2)}{N}, \ldots, 0\right]^T. \tag{13}$$

Based on the symmetry and assuming identical clamping branches with sufficiently large switching frequencies, the average clamping current of the $j^{\text{th}}$ clamping branch in the upper arm follows

$$i_{b_j,u}(t) = \max\left(\frac{1 + m_a \sin(\omega t)}{2} i_u(t) \Delta_a \frac{(N-j)j}{N-1}, 0\right), \tag{14}$$

where $j = 1, \dots, N-1$ and $i_u$ is the upper arm current. The max(◊) function shows the clamping branch conducts only in the positive direction. Similarly, the clamping current equation for the lower arm is

$$i_{b_j,l}(t) = \max\left(\frac{1-m_a \sin(\omega t)}{2} i_l(t)\Delta_a \frac{(N-j)j}{N-1}, 0\right). \quad (15)$$

Although symmetrical definition of the adjustments using (11) can help to reduce the effects of balancing in the load side, many of the balancing strategies do not follow such conventions. Nevertheless, the operating principles of the balancing algorithms remain unchanged.

Although the average value of the arm current is always positive, higher harmonic content, namely first-order and second-order components, can result in instances where the arm current is negative, which combined with the open–loop balancing techniques can reduce the balancing efficiency. Therefore, an improved level-adjustment control follows

$$\delta_{u,j} = \text{sgn}(i_{\text{arm},u})\Delta_a\left(\frac{1}{2} - \frac{j-1}{N-1}\right), \quad (16)$$

which has all the previous advantages in addition to improved efficiency.

Although no cell-sorting will be necessary, the level-adjustment according to (16) necessitates a closed-loop control, with two updates per fundamental cycle. Proposing a balancing method is not the main contribution of this paper, and the compensated model for diode-clamped MMCs in the next section is applicable to both open-loop and close-loop balancing techniques.

## IV. Proposed State-Space Model and Estimator

### A. Compensated State-Space Model

Inherent inter-dynamics of the system introduced by the clamping branch as well as the non-Gaussian distribution of the parameters (such as the capacitance) complicates the derivation of a complete model for a diode-clamped MMC to the point where only numerical solutions are viable. Additionally, the available analytical simplifications for conventional MMCs will not lead to very accurate results, mainly due to the effect of the balancing modulations as well as the clamping branch. Therefore, we extend the conventional model of the MMC to account for both clamping and balancing efforts, and then use an optimal estimator (e.g., Kalman filter) to estimate the voltage of each module [30].

The output voltage of the arm based on the module voltages and their states is

$$V_{arm} = \sum_{j=1}^{N} V_{c_j} S_{j1} = \boldsymbol{V_c^T} \times \boldsymbol{S_1}, \quad (17)$$

where $\boldsymbol{V_c}$ is the vector of module voltages in one arm and $\boldsymbol{S_1}$ is the vector of the upper switches in each module in the same arm with $S_{j1}$ representing its $j^{\text{th}}$ element.

The dynamics of the capacitor voltage of an inserted module follows

$$\dot{V}_{c_j} = \text{S}_{\text{j1}} \frac{i_{\text{arm}}}{C_j}, \quad (18)$$

where $C_j$ is the capacitance of the j$^{\text{th}}$ module in the arm. Applying the forward Euler discretization leads to

$$V_{c_j}^{(k)} = V_{c_\text{j}}^{(k-1)} + \dot{V}_{c_j}^{(k-1)} T_s, \quad (19)$$

where $T_s$ is the sampling period. Substituting (18) in (19) results in

$$V_{c_j}^{(k)} = V_{c_j}^{(k-1)} + \text{S}_{\text{j1}}^{(k-1)} \frac{T_s}{C_j} i_{\text{arm}}^{(k-1)}, \quad (20)$$

and the arm voltage can be written as

$$\boldsymbol{V_c^{(k)}} = \boldsymbol{V_c^{(k-1)}} + \mathbf{S_1^{\mathit{T}(k-1)}} \times \mathbf{I} \times \frac{T_s}{\mathbf{C}} i_{\text{arm}}^{(k-1)}, \quad (21)$$

where the vector $\boldsymbol{V_c^{(k)}} = \left[V_{c_1}^{(\text{k})}, V_{c_2}^{(\text{k})}, \dots, V_{c_N}^{(\text{k})}\right]^T$ is the states' vector, $\boldsymbol{S_1}(k-1)$ is a vector containing the switching signals of the modules ($\text{S}_{\text{j1}}(k-1)$), $\mathbf{I}$ is the identity matrix, and the vector $\frac{T_s}{\boldsymbol{C_j}}$ is a fixed vector which can be written as

$$\frac{T_s}{\boldsymbol{C_j}} = \left[\frac{T_s}{C_1}, \frac{T_s}{C_2}, \dots, \frac{T_s}{C_\text{N}}\right]^T, \quad (22)$$

and can be precalculated. Additionally, the $\frac{T_s}{\boldsymbol{C_j}}$ vector turns into a constant coefficient, if all the capacitances are identical.

Hence the average state-space model of the module voltages follows

$$\boldsymbol{V_c^{(k)}} = \mathbf{A_{ss}} \boldsymbol{V_c^{(k-1)}} + \mathbf{B_{ss}} i_{\text{arm}}^{(\text{k}-1)} + \boldsymbol{w}^{(k)}, \quad (23)$$

$$V_{\text{arm}}^{(k)} = \mathbf{C_{ss}} \boldsymbol{V_c^{(k)}} + v^{(k)}, \quad (24)$$

where the state matrices are $\boldsymbol{A_{ss}} = \mathbf{I}$, $\boldsymbol{B_{ss}} = \boldsymbol{S_1^{T(k-1)}} \times \mathbf{I} \times \frac{T_s}{C_u}$, $\boldsymbol{C_{ss}} = \mathbf{S_1^{(\mathit{k})}}$, and $\boldsymbol{D_{ss}} = 0$. Additionally, $\boldsymbol{w}(k)$ and $v(k)$ are the modeling and measurement noise, respectively [30].

In (22) and (23), the effect of the balancing efforts (e.g., level–adjustmet for [27], conduction time for [20], switching delays [16]) are considered, which leads to increasing the voltage of the lower modules in the arm and decreasing the upper ones. It is mainly bacause any balancing effort will directly affect the $\boldsymbol{S_1^{(k)}}$ vector. Nevertheless, without considering the effect of the clamping branch, the model is inadequate.

Two clamping branches connect each module to the previous (can only discharge said module) and the next modules (can only charge said module) except the first and last modules. Rewriting (19) while considering the effect of these two clamping branches, we arrive at

$$V_{c_j}^{(k)} = \left(1 - \frac{T_s}{2L_jC_j}\left(\begin{array}{c} b_j\left(1 - \text{S}_\text{j}^{(k-1)}\right) + \\ b_{(j+1)}\left(1 - \text{S}_{(\text{j}+1)}^{(k-1)}\right)\end{array}\right)\right) V_{c_j}^{(k-1)} + \frac{b_jT_s}{2L_jC_j}\left(1 - \text{S}_\text{j}^{(k-1)}\right) V_{c_{(j-1)}}^{(k-1)} + \frac{b_{(j+1)}T_s}{2L_jC_j}\left(1 - \text{S}_{(\text{j}+1)}^{(k-1)}\right) V_{c_{(j+1)}}^{(k-1)} + \frac{T_s}{C_j} \text{S}_j^{(k-1)} i_{\text{arm}}^{(k-1)}, \quad (25)$$

where $S_j = S_{j1}$, and $b_j$ is variable that acording to the switching frequency and modulation reference calculates the average time that the clamping path is forward-biased

$$b_j = \begin{cases} (1-m_a)T_{sw}, & V_{(j+1)} > V_j \\ 0, & V_{(j+1)} \le V_j \end{cases}. \quad (26)$$

Consequently, the compensated state-space model can be rewritten as

$$\boldsymbol{V_c^{(\text{k})}} = \mathbf{A'_{ss}} \boldsymbol{V_c^{(\text{k}-1)}} + \mathbf{B_{ss}} i_{\text{arm}}^{(\text{k}-1)} + \boldsymbol{w}^{(k)}, \quad (27)$$

$$v_{\text{arm}}^{(k)} = \mathbf{C_{ss}} \boldsymbol{V_c^{(k)}} + v^{(k)}. \quad (28)$$

where $a'_{pq} \in \boldsymbol{A'_{ss}}$ is the element in the $p^{\text{th}}$ row and the $q^{\text{th}}$ coloumn. The value of $a'_{pq}$ for all $q, p$ except the first and last module (i.e., $\forall p, q | p \ne 1, N$) follows

$$a'_{pq}=\begin{cases}\frac{b_pT_s}{2L_pC_p}\left(1-S_p^{(k-1)}\right), & q=p-1\\ 1-\frac{T_sb_p\left(1-S_p^{(k-1)}\right)}{2L_pC_p}-\frac{T_sb_{(p+1)}\left(1-S_{(p+1)}^{(k-1)}\right)}{2L_pC_p}, & q=p\\ \frac{b_{(p+1)}T_s}{2L_pC_p}\left(1-S_{(p+1)}^{(k-1)}\right), & q=p+1\\ 0, & \text{else}\end{cases}. \quad (29)$$

If $p=1$, then $\forall q\in(1,\dots,N)$

$$a'_{1q}=\begin{cases}1-\frac{T_sb_{(p+1)}}{2L_qC_q}\left(1-S_{(p+1)}^{(k-1)}\right), & q=1\\ \frac{T_sb_{(p+1)}}{2L_pC_p}\left(1-S_{(p+1)}^{(k-1)}\right), & q=2\\ 0, & \text{else}\end{cases}, \quad (29)$$

and for $p=N$ and $\forall q\in(1,\dots,N)$, $a'_{Nq}$ follows

$$a'_{Nq}=\begin{cases}\frac{T_sb_p}{2L_pC_p}\left(1-S_p^{(k-1)}\right), & q=N-1\\ 1-\frac{T_sb_p}{2L_pC_p}\left(1-S_p^{(k-1)}\right), & q=N\\ 0, & \text{else}\end{cases}. \quad (31)$$

If the sampling frequency is low or due to asynchronized sampling, the small balancing efforts can be overlooked and and lead to an increased error.

Averaging the modulation index of each module in one cycle of the output voltage results in

$$\overline{m}_{u,j}=\frac{1}{2}-\delta_{u,j}, \quad (32)$$

which ideally should reflect the average value of the $S_{1j}(t)$ in one cycle. But, due to low sampling frequencies, high delays in the measurements, and/or asynchronized sampling (i.e., one state of a module is observed more than the other state), it is possible that by average of the discrete values of $S_{1j}(k)$ per

$$\bar{S}_{j1}=\frac{1}{f_s}\sum_{z=k}^{k-f_s}S_{j1}(z), \quad (33)$$

is not equal to $\overline{m}_{u,j}$, that means the balancing effort is not calculated correctly in the model. Therefore, in order to compensate the negative effect of sampling, an addional term can be added to the model by defining a compensated state vector ($\boldsymbol{S}'_1$) according to

$$\boldsymbol{S}'_1(k)=\boldsymbol{S}_1(k)-(\bar{\boldsymbol{S}}_1-\overline{\boldsymbol{m}}_u). \quad (34)$$

In (34), $\bar{\boldsymbol{S}}_1$ is the vector of the averaged switch states in one fundamental cylce, and it is constantly updated through a sliding window at each iteration. $\overline{\boldsymbol{m}}_u$ is the vector of the expected switch states which can be easily calculated for each module per (9). Substituting $\boldsymbol{S}'_1(k)$ in (22) and (23) result in a model that can consider the balancing efforts more accurately.

### B. Optimal Estimation Algorithm

In the last step, an optimal estimation algorithm is integrated with the proposed state-space model to estimate the voltages of the modules [31]. We select Kalman filter as a prime example of an optimal estimator to verify the performance of the developed model. Table I lists the pseudo-code of the KF algorithm used to estimate the module voltages. However, other state-of-the-art estimators can be accordingly implemented with the proposed model, and KF is only an example used for a quantitative comparison between the conventional and compensated models.

Table I: Kalman filter for estimating the module voltages of the MMC

| **STEPS** | | |
|---|---|---|
| **STEP 0** | Initialization | $\hat{\boldsymbol{V}}_c^{(k-1)},\mathbf{P}^{(0)+},\boldsymbol{Q},R$ |
| **STEP 1** | Projection | $\boldsymbol{V}_c^{(k)-}=\mathbf{A}'_{ss}\boldsymbol{V}_c^{(k-1)+}+\mathbf{B}_{ss}i_{arm}^{(k-1)}$ <br> $\boldsymbol{P}^{(k)-}=\mathbf{P}^{(k-1)+}+\boldsymbol{Q}$ |
| **STEP 2** | Calculate KF gain | $\boldsymbol{K}_g^{(k)}=\frac{\boldsymbol{P}^{(k)-}\mathbf{C}_{ss}{}^T}{\mathbf{C}_{ss}\boldsymbol{P}^{(k)-}\mathbf{C}_{ss}{}^T+R}$ |
| **STEP 3** | Correction | $\boldsymbol{V}_c^{(k)+}=\boldsymbol{V}_c^{(k)-}+\boldsymbol{K}_g^{(k)}\left(v_{arm}^{(k)}-\mathbf{C}_{ss}\boldsymbol{V}_c^{(k)-}\right)$ <br> $\boldsymbol{P}^{(k)+}=\left(\boldsymbol{I}-\boldsymbol{K}_g^{(k)}\mathbf{C}_{ss}\right)\boldsymbol{P}_k{}^-$ |
| **STEP 4** | Return to **STEP 1** | $\boldsymbol{V}_c^{(k)+},\boldsymbol{P}^{(k)+}\Rightarrow\boldsymbol{V}_c^{(k)+},\boldsymbol{P}^{(k)+}$ |

In most power electronics systems, the measurement noise as well as the parameter imbalances do not have any type of uniform distribution; therefore, we ultimately should heuristically define the initial covariance matrix ($\boldsymbol{P}$) for the KF [32]. The measurement noise ($R$) and modeling error ($\boldsymbol{Q}$) are usually assumed to be independent for each module as well as independent from each other [33]. Therefore, $\boldsymbol{Q}$ is defined as a diagonal matrix, and $R$ as a scaler value, where the initial values should again be set heuristically. However, knowing the effect of each parameter on the estimator's behavior can be of benefit in the beginning of our search. Increasing the value of $\boldsymbol{Q}$ leads to more a rapid convergence and prevention of divergence. However, high values of $\boldsymbol{Q}$ will increase fluctuations. On the other hand, reducing the value of $\boldsymbol{Q}$ will reduce the estimation ripples, but also makes the convergence slower and increases the error. Sensitivity to noise can be reduced by increasing $R$ at the cost of slower dynamics.

Comparing the developed model to the conventional one shows that $\boldsymbol{A}'_{SS}$ is not diagonal anymore and has the general form as shown in (35). However, forming the complete $\boldsymbol{A}'_{SS}$ for STEP 1 to perform matrix multiplications is unnecessary. The projected voltage of each module in STEP 1 can be calculated using simple scalar mathematics starting from the first module.

$$\boldsymbol{A}'_{SS}=\begin{bmatrix} a'_{11} & a'_{11} & & 0 & & 0\\ a'_{21} & a'_{22} & a'_{23} & & \cdots & \\ & a'_{32} & a'_{33} & a'_{34} & & \\ 0 & & & \ddots & & 0\\ & \vdots & & & \ddots & \\ 0 & & 0 & & a'_{N(N-1)} & a'_{NN}\end{bmatrix} \quad (35)$$

In the proposed approach, the estimator is formulated such that each state equation contains a maximum of three state variables and the whole state-space model contains only one output variable. Also, the computations regarding the output equation are entirely simple scalar calculations, which makes the computations less demanding for an online implementation in large systems.

### B. Discussion

It is possible to operate the diode-clamped MMC topology without any direct measurements at module level to reduce the cost and complexity of the system. However, since the system is running open-loop, the required variations in the duration of the serial connections must be designed for a worst-case scenario, which may not be an optimum solution regarding added losses. Additionally, in many applications, monitoring and protection requirements would still necessitate tracking of all the modules' voltages.

Table II: General comparison of the state-of-the-art estimators in literature

| Method | Voltage Sensors | Advantages and Disadvantages |
|---|---|---|
| Based on PSO* for conventional MMCs [34] | 2 | − Computationally extreme; − relatively low accuracy; −very slow convergence speed; −poor performance under rapid variations in operating point |
| Based on PSO considering parallel modes [35] | 2 | − Computationally extreme; −very slow convergence speed; −poor performance under rapid variations in operating point;+high accuracies |
| Based on two step estimation (model and measurement) [36] | 2 | − Affected by the sampling delay; − neglects the resistance effect; + computationaly moderate |
| Based on energy variations in the capacitor [37] | 2 | − Affected by sampling delay; − neglects the resistance effect; −slow convergence; +computationally moderate |
| Iterative based method with delay compensation [22] | ≥ 4 | − Neglects the resistance effect; − needs at least two extra voltage sensors; −computationaly demanding; + more stable; +considers the sensor behavior on measured signal |
| Voltage estimation based on Adaline iterative algorithm [23] | 2 | − Affected by sampling delay; − neglects the resistance effect; − Adaline method convergence speed is lower than KF; −lower noise sensitivity;+ computationaly moderate |
| Based on exponentially weighted recursive least square [38] | ≥ 2 | − Neglects the resistance effect; −neglects the clamping effect; −neglects the balancing effect; −computationally extreme |
| Based on discrete-time sliding mode observer (DTSMO) [39] | 2 | − Neglects the resistance effect; −neglects the clamping effect; −neglects the balancing effect; −moderate convergance speed; +computationally moderate |
| Based on KF for conventional MMCs [31] | 2 | − Affected by sampling delay; − neglects the resistance effect; −neglects the clamping effect; −neglects the balancing effect; +computationally moderate; +good noise sensitivity |
| EKF for conventional MMCs [40] | 2 | − Affected by sampling delay; −neglects the clamping effect; −neglects the balancing effect; −computationally demanding; +estimates the arm currents (down not need arm current sensors); +good noise sensitivity; +suitable convergance speed |
| KF with grouping measurement [41] | ≥ 4 | −May need higher number of sensors; −computationally extreme− neglects the resistance effect; −neglects the clamping effect; −neglects the balancing effect;+ considers capacitor effect; +more robust; +estimates capacitor value |
| Based on Dual KF for conventional MMCs [26] | 2 | −Neglects the clamping effect; −neglects the balancing effect; −computationally more demanding; −can be slow in tracking fast variation in capacitor voltage; + low number of required sensors; + Estimates the effective resistance of each module |
| **Proposed method** | 2 | −Computationally demanding; −neglects the capacitor variation over time (future work);+ low number of required sensors; +considers the clamping effect; +considers the balancing effect; +very good accuracy in imbalanced/unstable conditions; +stable with low-frequency switching methods; +compensates for the sampling delay or low-frequency sampling |

* PSO: Particle Swarm Optimization; ** KF: Kalman Filter; *** EKF: Extended Kalman Filter

Many estimation techniques that can track the voltage of each module are available for conventional MMCs [1, 24]. However, since they do not consider the effect of the clamping branch, none are applicable to a diode-clamped MMC without significant errors. Additionally, none considers the effect of the balancing algorithms. Table II offers an overview of the state-of-the-art in comparison with the proposed technique. The main advantages are considering the balancing effect as well as the clamping branch influence on the voltages, proposing a technique to compensate the effect of sampling, which is neglected in other methods, and superior performance in case of imbalance. The proposed method requires only two voltage sensors per phase (6 for three phases) instead of one per module ($6N$ for three phases), which is on par with other state-of-the-art methods, and compared to the conventional methods, it can reduce the number of sensors from $2N$ to only 2. Additionally, while we use KF as an example of an optimal estimator, the developed model may be also used in combination with other state-of-the-art estimators.

Table III: Parameters for Simulation and Experiments

| Parameters | | Simulation | Experimental |
|---|---|---|---|
| Rated Power | | $P_{ac} = 1.14$ MW | $P_{ac} = 2$ kW |
| Load Inductor | | $L_L = 0.1$ mH | $L_L = 0.1$ mH |
| Module rated voltage | | $V_{sm} = 1.2$ kV | $V_{sm} = 45$ V |
| DC link voltage | | $V_{dc} = 9.6$ kV | $V_{dc} = 230$ V |
| Number of modules | | 16 | 10 |
| Arm inductance | | $L_{arm} = 5$ mH | $L_{arm} = 2$ mH |
| Arm Resistor | | $R_{larm} = 50$ mΩ | $R_{larm} = 20$ mΩ |
| Carrier frequency | | $f_c = 2$ kHz | $f_c = 2$ kHz |
| Sampling frequency | | $f_s = 10$ kHz | $f_s = 10$ kHz |
| Output frequency | | $f = 50$ Hz | $f = 50$ Hz |
| Modules | Capacitance | $C_{cap} = 6$ mF | $C_{cap} = 2.5$ mF |
| | Capacitor resistor | $R_{cap} = 2$ mΩ | $R_{cap} = 2$ mΩ |
| Modulation index | | $m_a = 0.5$~$0.95$ | $m = 0.9$ |
| Clamping circuit | Inductance | $L_{ld} = 10$ µH | $L_{ld} = 1$ µH |
| | Resistor | $R_{ld} = 0.5$ mΩ | $R_{ld} = 0.5$ mΩ |

## V. SIMULATION AND EXPERIMENTAL RESULTS

A single-phase model with $2 \times 8 = 16$ modules ($N = 8$) and 14 clamping branches verifies the feasibility of the proposed model and estimator, and studies the general behavior of the system. Additionally, a scaled-down prototype provides proof of concept in an experimental setup. Table III lists the parameters of the simulation and the experimental systems. The specfcation of the power modules SEMiX854GB176HDs from Semikron are used for the semiconductors. The switching frequency of each module is 2 kHz and the amplitude of the modulation reference varies in the range of 0.50 to 0.95. Additionally, the sampling frequency of the KF and the sampling rate for the simulations and the experimental setup are 10 kHz.

### *A. Simulation Studies*

We study the behavior of the estimators under balanced conditions with identical modules and imbalanced conditions with mismatch between the module capacitances and self-discharging rates. In the second case, a normal distribution is considered for the capacitance and resistance of the modules with $\pm 15\%$ spread from the rated values, with the first module having the lowest capacitance and internal resistance. Additionally, the self-discharges of Modules 2, 4, 7, and 8 are increased with Module 8 having the highest discharge to emulate a worst-case scenario.

Fig. 5(a) and (b) show the phase voltage and current waveforms for balanced and imbalanced systems with $\Delta = 0$ and $\Delta = 0.02$, respectively, which verifies that the output is not affected by the balancing effort. Additionally, Fig. 6(a) and (b) show the results of the Kalman filters based on the developed state-space model as well as the conventional model of MMCs. In Fig. 6(a), yellow color marks the true voltages, blue depicts the results using the conventional MMC model, and red depicts the results from the proposed method. The results confirm that both models are stable. However, while both models converge, the modified state-space model can achieve slightly lower absolute errors. According to Fig. 6(b), the maximum error for the proposed estimator in an ideal system is below 0.5%.

Fig. 7 shows the behavior of the estimator in a heavily imbalanced system with $\Delta_a = 0$. Due to the imbalance and without any balancing effort, the voltages start to diverge. Fig. 7(a) illustrates measured and estimated voltages. Similar to the previous case, both of the estimators converge, but the proposed model tracks the measured voltage more closely. Fig. 7(b) verifies a >50% improvement in the estimation error compared to the conventional model of the MMC as well as a reduction in the amplitude of the error fluctuations.

The third scenario investigates the behavior of the estimators in the imbalanced system with non-zero level-adjustment ($\Delta_a = 0.02$). Fig. 8(a) presents the profile of the measured and estimated voltages of the capacitors in one arm. Similar to the previous scenarios, the KF based on the developed model can track the capacitor voltages with better accuracy. Furthermore, as seen in Fig. 8(b), despite the mismatches in the capacitances as well as the internal discharge rates, all the voltages converge to the rated value. According to Fig. 8, the maximum instantaneous error is <9 V, which is 40% lower compared to the conventional one.

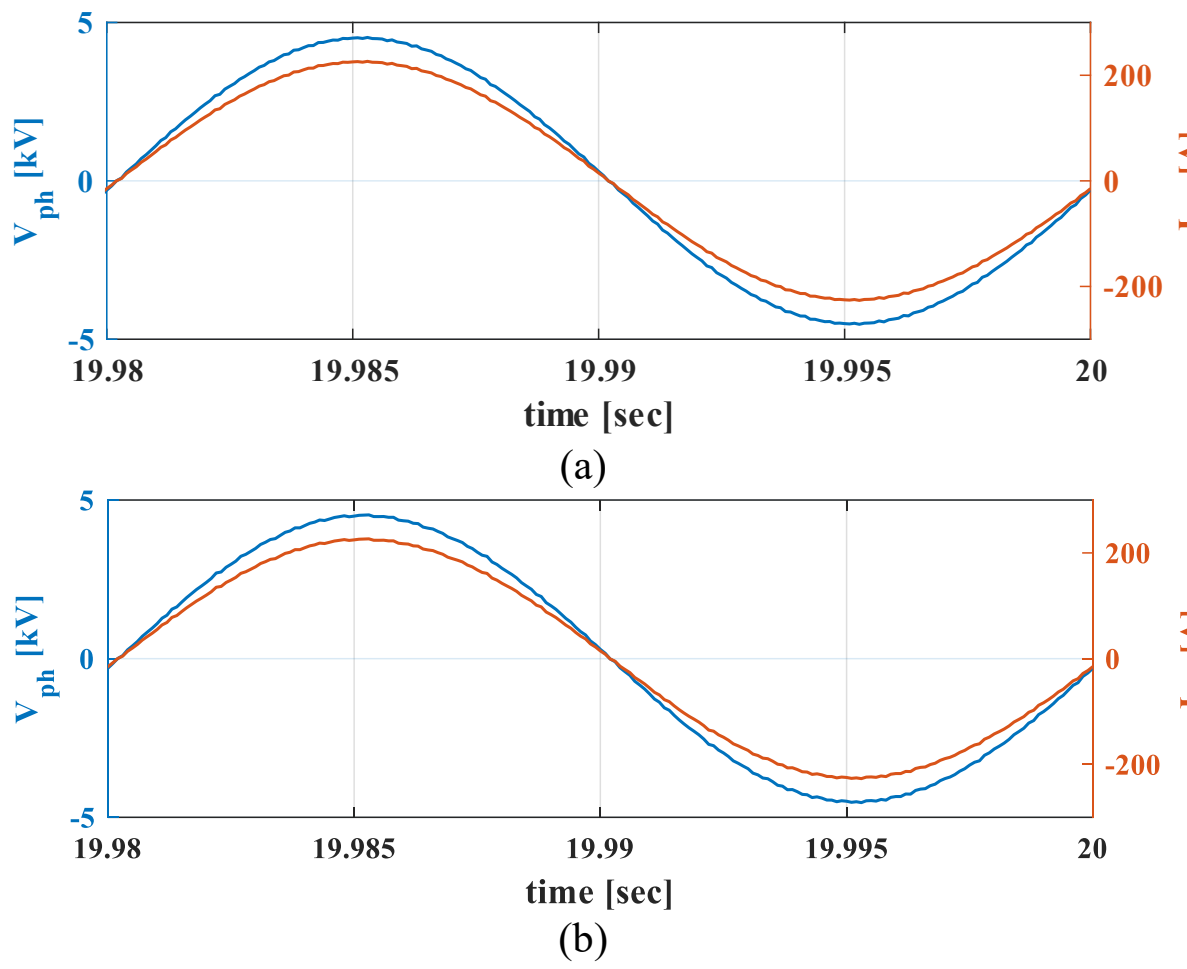

Fig. 5. Output phase voltage and current waveform: (a) balanced system with $\Delta_a = 0$; (b) imbalanced system with $\Delta_a = 0.02$

Since normally the switching frequency is low in larger systems, Fig. 9 compares the accuracy of the estimators with low-frequency modulation ($f_{sw} = 200$ Hz) as well as step variations in the modulation index. The result confirm that the proposed model can easily track the voltage with sufficient accuracy. Lower switching frequencies can lead to slightly higher absolute errors; however, the results are still well below the 1% mark. Additionally, lower modulation indices can increase the average duration that the clamping-branch is conducting. Therefore, the compensated model can offer better accuracy, and slightly reduce the amplitude of the fluctuations.

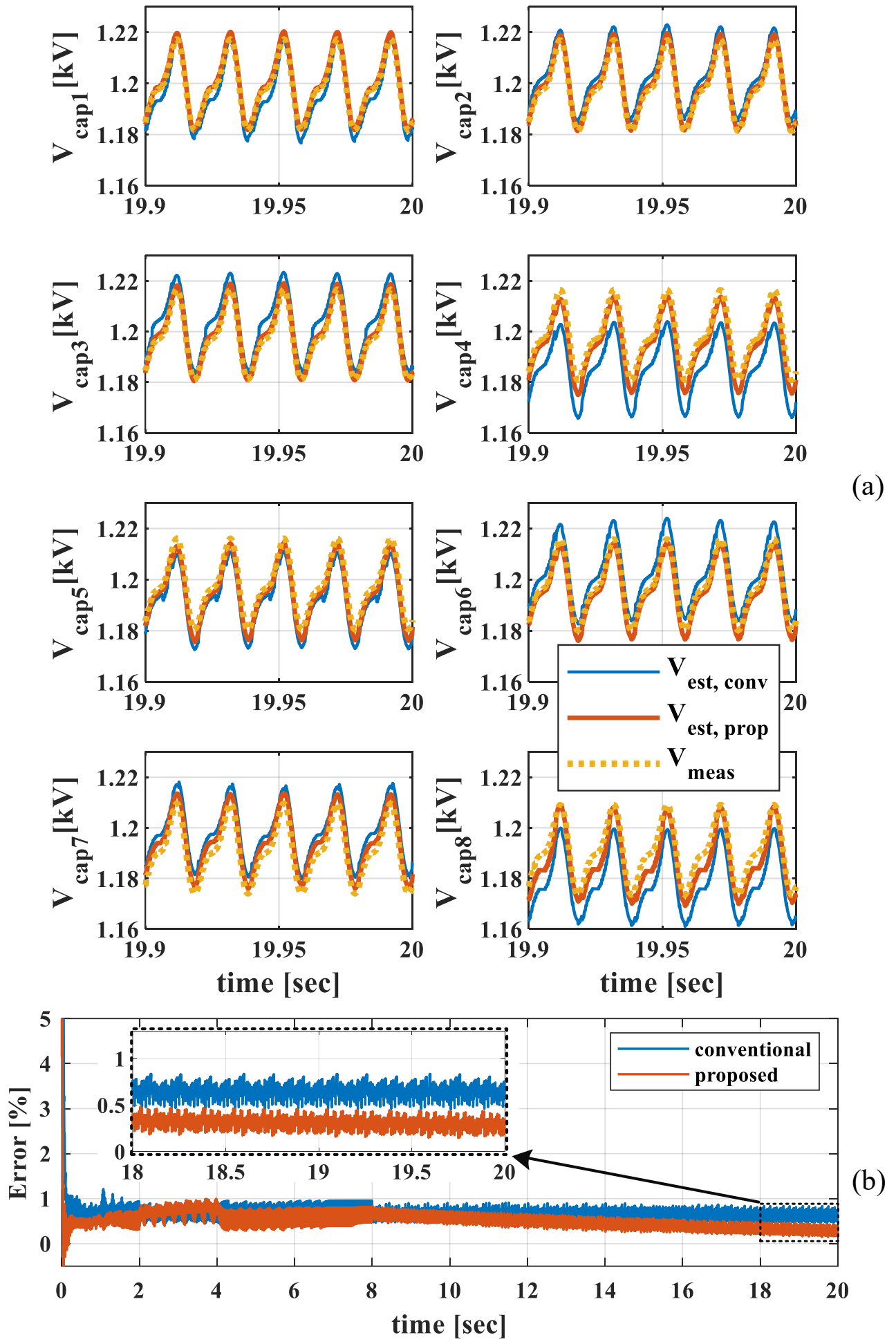

Fig. 6. Simulation results for a balanced system: (a) estimation of the capacitor voltages for conventional and proposed method; (b) profile of the maximum estimation error

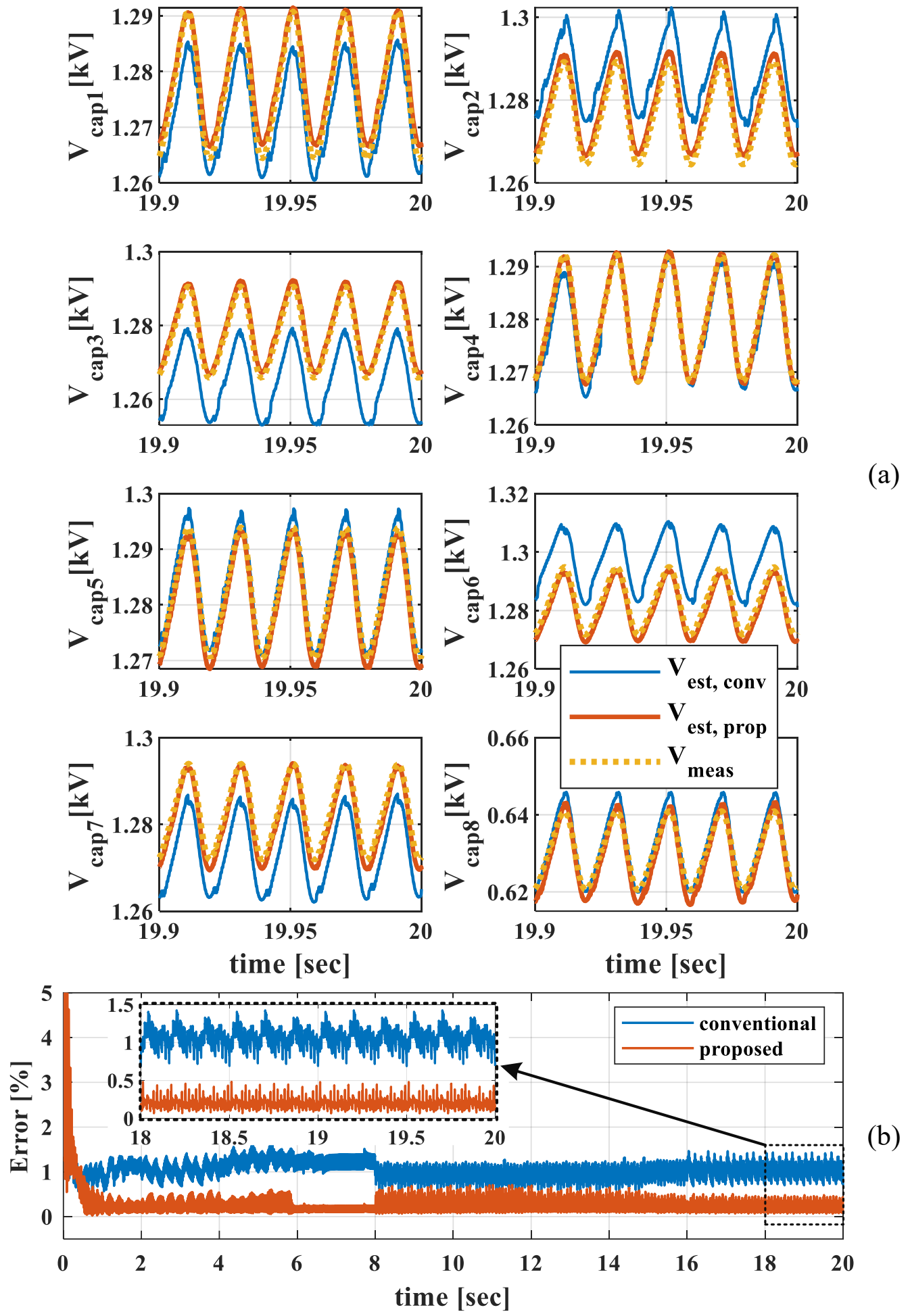

Fig. 7. Simulation results for an imbalanced system without balancing ($\Delta_a = 0$): (a) estimation of the capacitor voltages for conventional and proposed method; (b) profile of the maximum estimation error

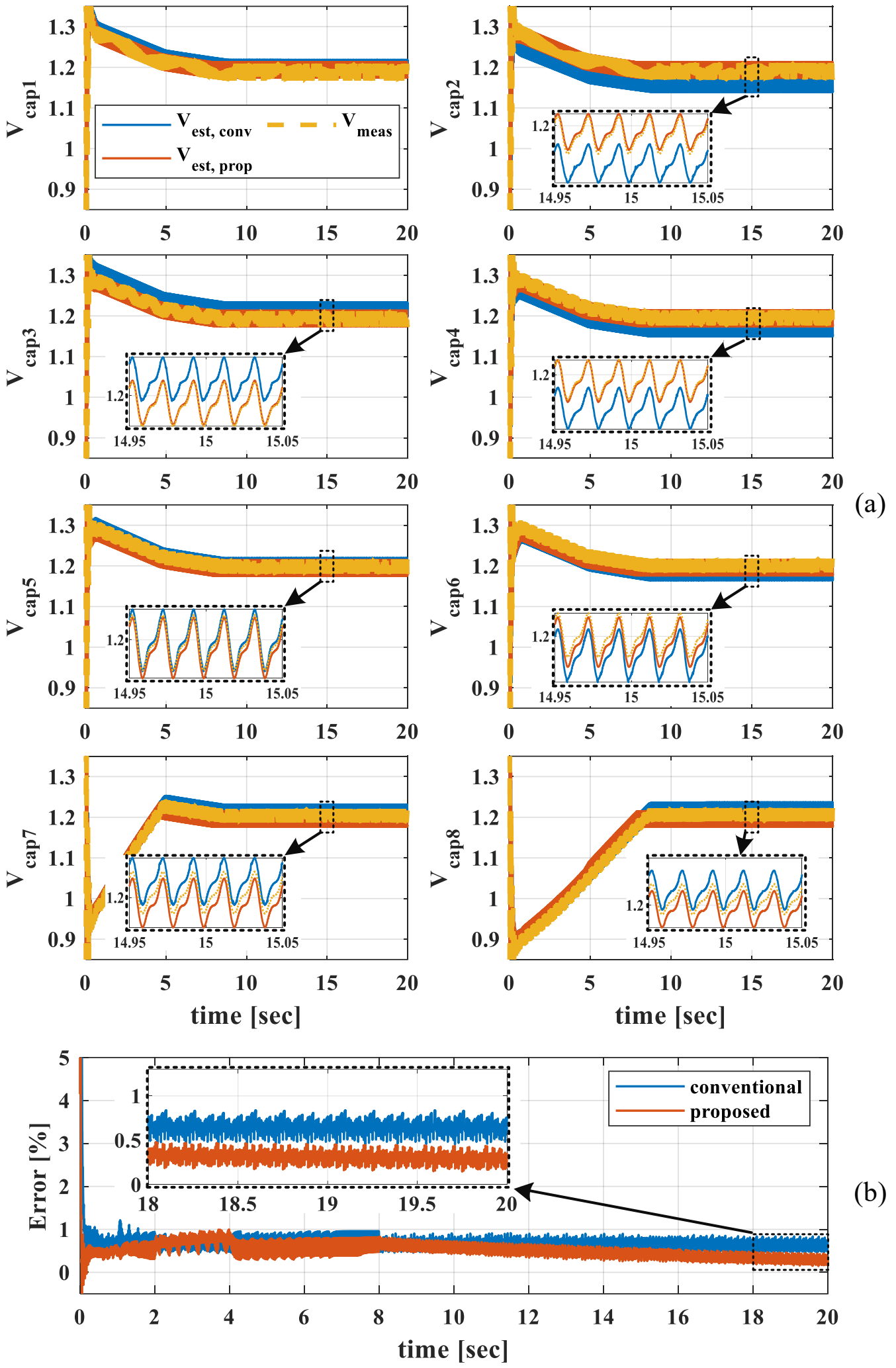

Fig. 8. Simulation results for an imbalanced system with $\Delta_a = 0.02$: (a) estimation of the capacitor voltages for conventional and proposed method; (b) profile of the maximum estimation error

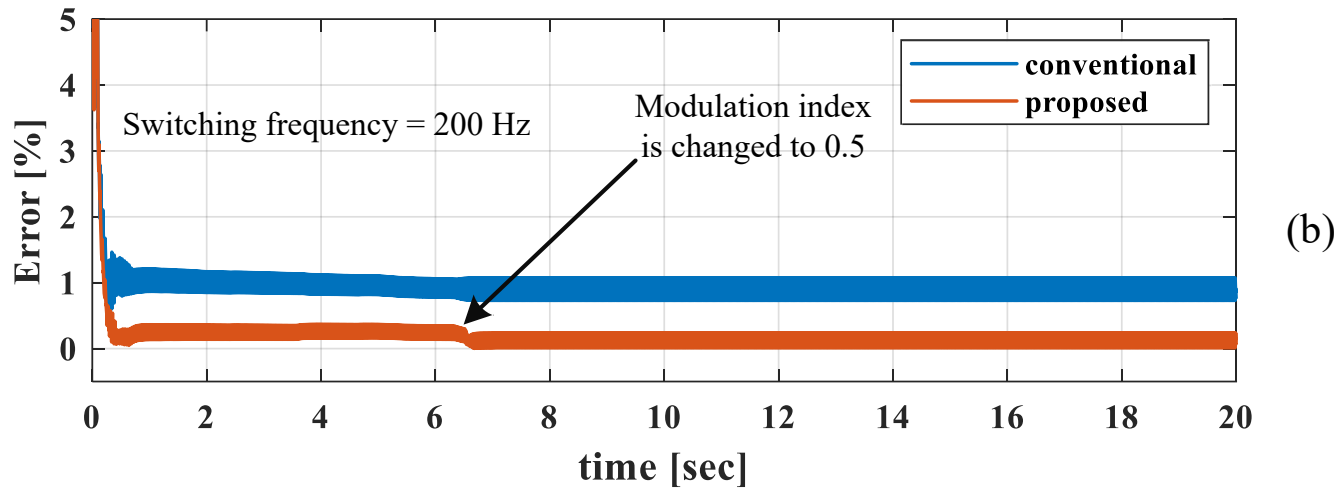

Fig. 9. Profile of maximum error with $\Delta_a = 0$ and low-frequency modulation

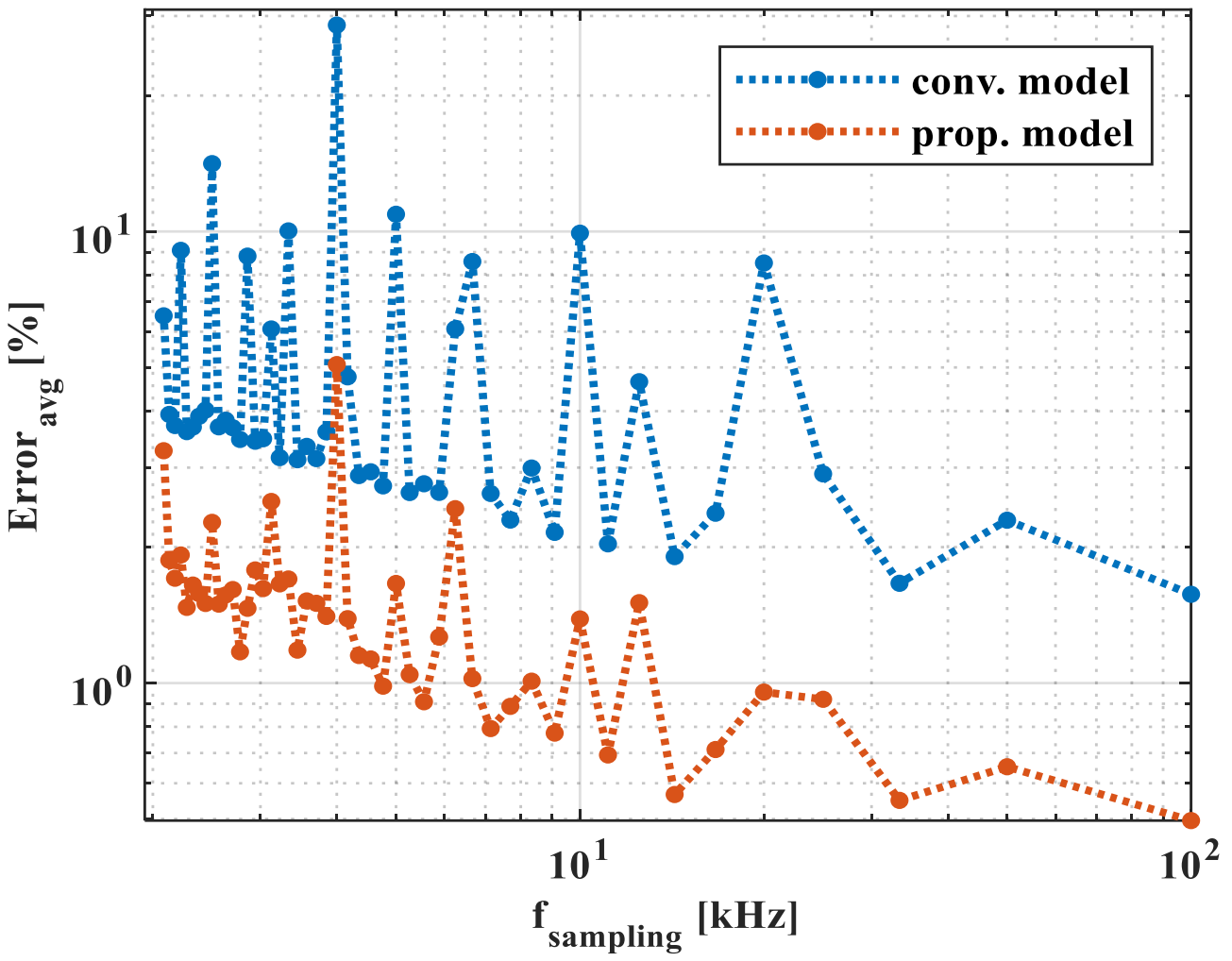

Fig. 10. Profile of the average error with respect to the sampling frequency

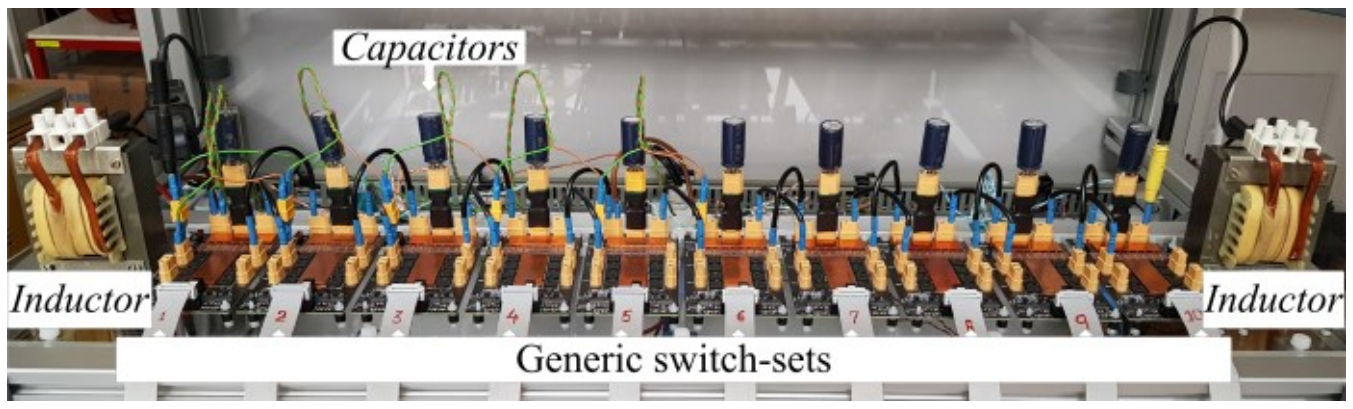

Fig. 11. Picture of the diode-clamped MMC prototype

Lastly, we compare the performance of the proposed model as well as the conventional model to the sampling frequency in Fig. 10. With reducing the sampling frequency, the estimation error increases. However, due to the developed sampling compensation technique, the proposed model can achieve better results.

As seen, in all scenarios the proposed model outperforms the conventional one with 30% to 50% error reduction. Additionally, the proposed method is widely unaffected by level-adjustment, lower sampling frequencies, or modulation index variations.

## B. Experiments

A scaled-down diode-clamped MMC verifies the estimator's performance, and Fig. 11 shows a photo of the setup. Each module includes about 500 µF low-ESR ceramic capacitors in parallel with 2.2 mF electrolytic ones. Labview in combination with an FPGA development board (National Instruments sbRIO 9627) perform the estimation and control functions.

Similar to the simulations, Fig. 12 (a) shows the estimated and measured arm voltages as well as the estimation error for $\Delta_a = 0$. The maximum error after removing the switching noise from the measurements is 4 V, which corresponds to less than 2% and verifies the capability of the proposed method. Additionally, Fig. 12(b) depicts the measured and estimated capacitor voltages in addition to the estimation error. The maximum estimation error is below 3.5%. Since no balancing method is implemented in this case, the capacitor voltages do not completely balance, even though the clamping branch ensures that $V_{c1} \geq V_{c2} \geq V_{c3} \geq V_{c4} \geq V_{c5}$.

The balancing behavior as well as the performance of the estimator with non-zero circulating current for $\Delta_a = 0.02$ are shown in Fig. 13. With $\Delta_a = 0.02$, the voltages converge to 45 V and the system has a stable operation after reaching its steady state condition. The maximum estimation error in this case is also below 2%.

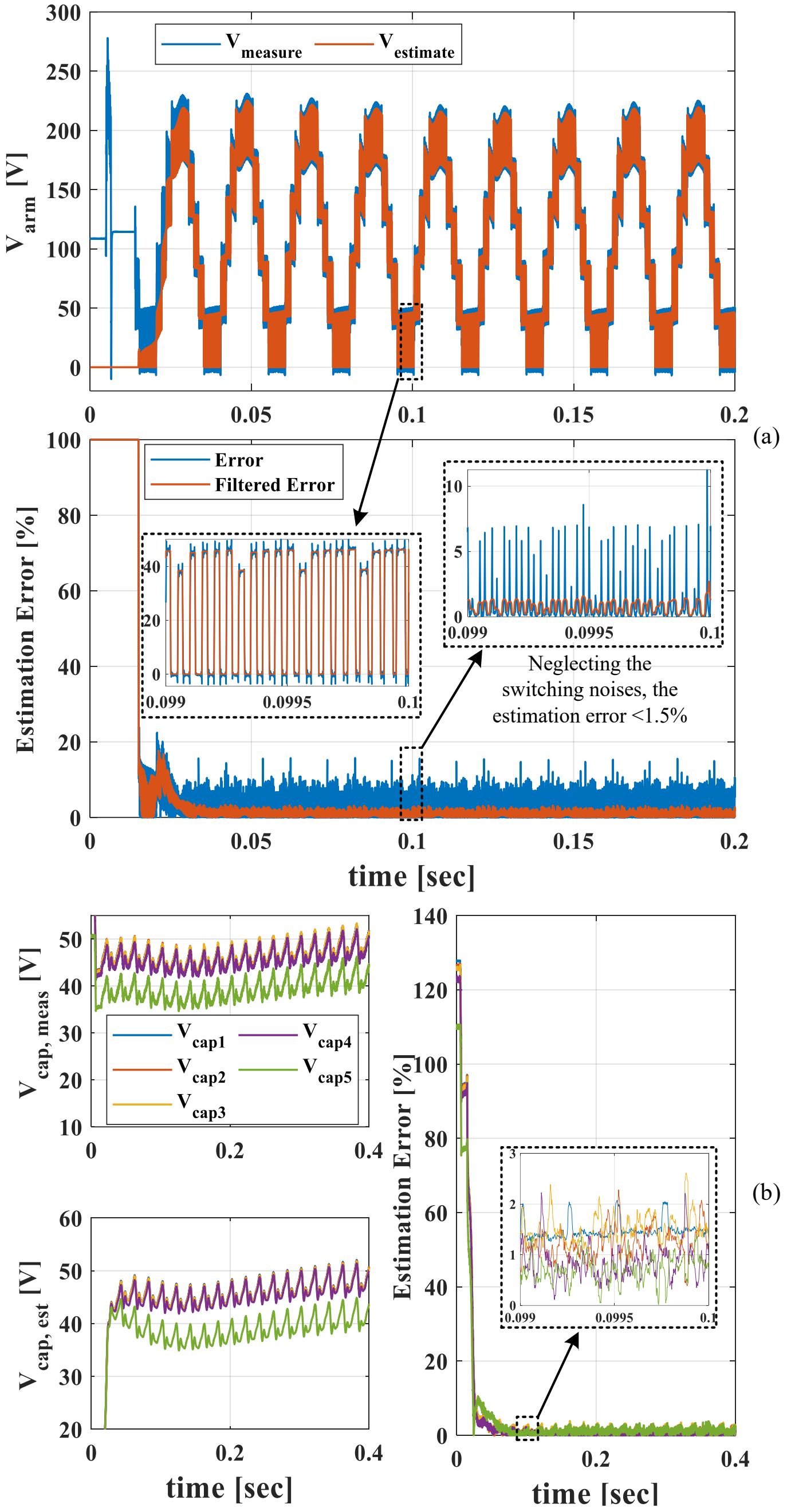

Fig. 12. Results for $\Delta_a = 0$: (a) measured and estimated arm voltage, as well as error of the arm voltage estimation; (b) capacitor voltages and estimation results as well as the maximum estimation error

## VI. Conclusion

This paper proposes a state-space model for diode-clamped MMCs, which is applicable for the available sensorless balancing techniques for this topolgy. Additionally, it provides a procedure to use an optimal estimator to achieve sensorless operation at the module levels, while still fully tracking the module voltages. The provided analysis as well as the simulation and experimental results verify the superiority of the developed model compared to the conventional MMC model.

The comaprison in Table II clearly shows significant advantages over the state-of-the-art, and comparison of the results from the proposed model with the ones from the conventional MMC show 30% to 50% reduction in estimation error. The improvement is even more noticable in imbalanced systems and/or with larger balancing currents.

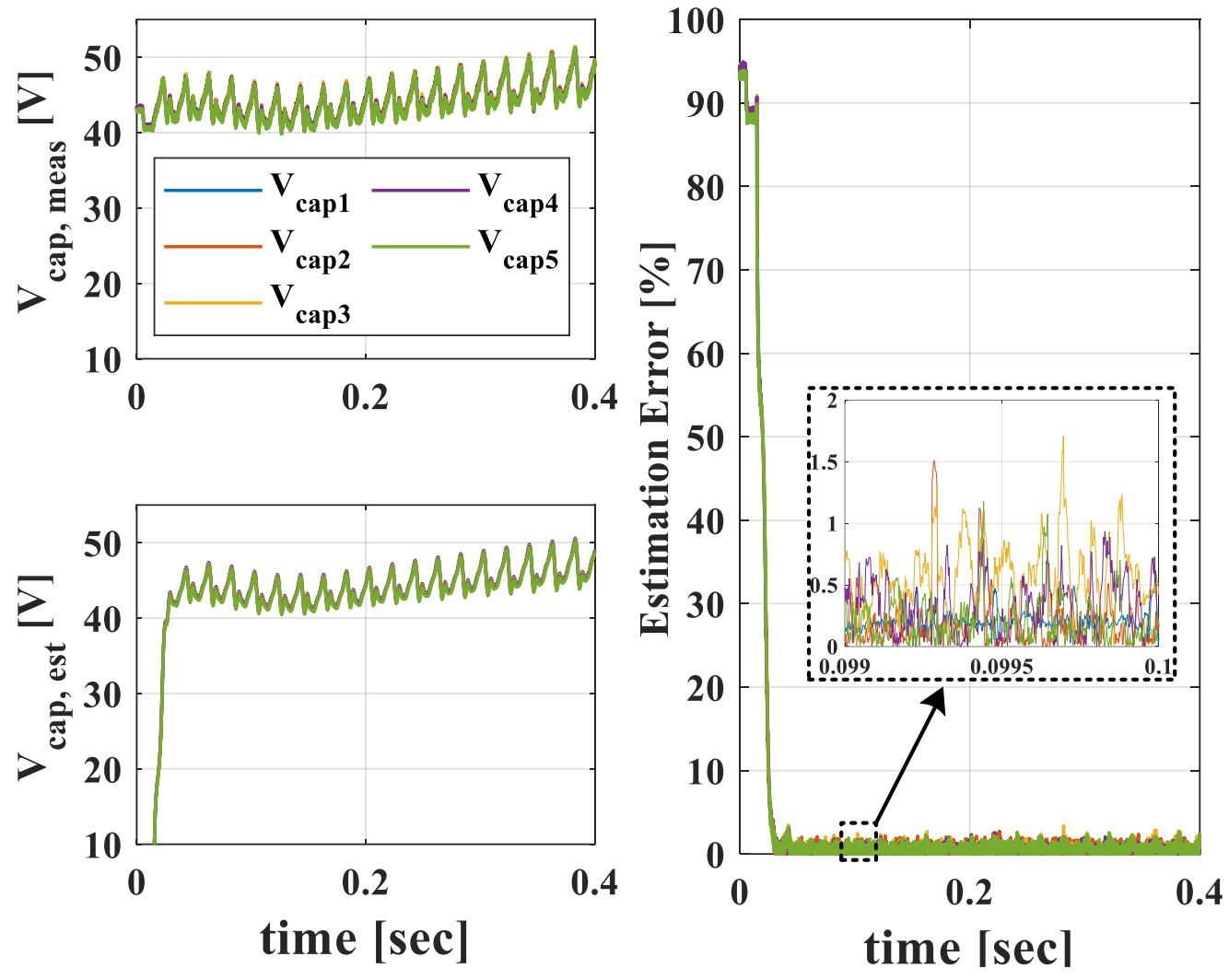

Fig. 13. Estimation results of imbalanced system with $\Delta_a = 0.02$

The low sampling compensation technique can allow for further reduction of the sampling frequency as well as better sensitivity to the small balancing efforts that might have been neglected in the conventional models with lower sampling frequencies.

Based on the experiments, the estimator can achive accuracies $\geq 97.5\%$ at all times with a wide range of $\Delta_a$=[0~0.04] and considering a 15% tolerance in the capacitances of the modules. Additonally, in a single-phase system, the estimator can reduce the number of voltage sensors from $2N$ (one voltage sensor per module) to only two sensors (arm voltage sensors). Although a closed-loop balancer is not the main focus of this paper, the proposed model and estimator can be used to achieve a more optimized closed-loop balancing in diode-clamped MMCs without any direct measurement at the module level, and it can minimize the balancing losses in the system.